\documentclass[prd,nofootinbib,tightenlines,twocolumn%,preprint%twocolumn%
]{revtex4}
\usepackage{soul,todonotes}
\usepackage[utf8]{inputenc}
\usepackage{enumitem} % Includes lists
\usepackage{mathtools}
\usepackage{amsmath}
\usepackage{braket}
\usepackage{amsfonts}
\usepackage{graphicx}

\newcommand{\intno}{\int_{0}^\infty}
\newcommand{\suminf}{\sum_{k=0}^\infty}

\newcommand{\NN}{{\mathbb{N}}}

\begin{document}

\title{Explicit examples of probability distributions for the energy density in two-dimensional conformal field theory}
\author{Matthew C. Anthony}
\author{Christopher J. Fewster}\email{chris.fewster@york.ac.uk}
\affiliation{Department of Mathematics, University of York, Heslington, York YO10 5DD. U.K.}
\date{\today}

\begin{abstract}
    Measurements of a weighted energy density average taken in the vacuum state of a conformal field theory in $1+1$ dimensions are randomly distributed with vanishing expectation value. The probability distribution is computed in closed form for two infinite families of averaging functions, generalising previously known examples. These examples may be further generalised by restriction to a half-line In all cases the distribution is that of a shifted Gamma distribution. 
\end{abstract}

\maketitle

\section{Introduction}

Repeated measurements of a quantum observable produce results that are statistically distributed in a manner determined by the observable and the state of the system. For example, if the observable is a local weighted average of the energy density of a quantum field, measured in the vacuum state, then then the distribution has vanishing expectation value, but there are nonzero probabilities for both positive and negative measurement outcomes due to quantum fluctuations. Knowledge of the distribution therefore provides information about these fluctuations and the extent to which they may outweigh thermal fluctuations in certain cases, with potentially observable consequences, see e.g.,~\cite{HuangFord:2017}. 

There has been recent progress in understanding the probability distribution for measurements of local averages of the energy density or related quantities, such as Wick squares of various operators \cite{FewFordRom,FewsFord:2015,FewForRom:2012,SchiaFewsFord:2018,FewHol2} or the `full counting statistics' of energy transfers in nonequilibrium  situations~\cite{GawedKozl:2019}. In particular, for conformal quantum field theory (CFT) in $1+1$ dimensions, it has been possible to determine the distribution in closed form for various averages of the stress energy tensor~\cite{FewFordRom, FewHol2} in the vacuum and certain other states, including thermal and highest weight states~\cite{FewHol2}.
Two methods have been developed, both of which rely on the solution to a nontrivial subsidiary problem. The first calculates the moment generating function using the solution to a nonlinear integro-differential equation based ultimately on the Ward identities~\cite{Haba,FewFordRom, FewHol2}. The second computes the Fourier transform of the probability distribution using the solution to a conformal welding problem~\cite{FewHol2}. The welding method of~\cite{FewHol2} has been extended to the full counting statistics problem in~\cite{GawedKozl:2019}; here, instead of welding two discs to produce a Riemann sphere, as in~\cite{FewHol2}, one must weld the boundaries of an annulus to produce a torus.

Although both methods are general in scope, relatively few closed form examples are yet known. The only example treated in~\cite{FewFordRom} was Gaussian averaging; this remained the only known example until quite recently, when three infinite families of examples were discovered~\cite{FewHol2}. 
In this note we document some further examples in which closed form results may be obtained using the moment generating function approach. To be precise, let $T(u)$ be a chiral component of the stress energy tensor of a unitary positive energy CFT in $1+1$ dimensions, with central charge $c$. (Relevant background on CFT may be found in~\cite{FewHol1,FewHol2} and references therein.) Then we will determine the probability distribution for measurements of the local average
\begin{equation}
    T(f):= \int_{-\infty}^\infty T(u) f(u)\,du
\end{equation}
made in the vacuum state, for test functions $f$ belonging to two infinite families. One family is based on the Gaussian function
\begin{equation}\label{eq:mGaussian}
f_{a,b}(u) =  \gamma u^{2a} e^{-bu^2},\qquad 
    a\in \mathbb{N}_0, \quad b>0,
\end{equation}
where $\gamma$ is a normalisation factor, while the other is based on the Lorentzian function
\begin{equation}\label{eq:g}
    g_{n,a,b}(u) = \frac{Cu^{2a}}{(b^2+u^2)^n}, \quad a,n\in\mathbb{N}_0,~0\le a< n,~b>0,
\end{equation}
where $C$ is again a normalisation constant.
The case $a=0$ of each family is known from previous work~\cite{FewFordRom, FewHol2}. We also study half-line variants of these test functions in the cases $a\ge 1$. Remarkably, in all these cases, the probability distribution takes the same general form, namely a shifted Gamma distribution with the probability density function
\begin{equation}\label{eq:shiftedGamma}
    P(\omega) = \Theta (\omega + \sigma) \frac{ \beta ^ {\alpha} (\omega+\sigma)^{\alpha-1}}{\Gamma                      (\alpha)}e^{-\beta(\omega+\sigma)},
\end{equation} 
where $\Theta$ is the Heaviside function, and the parameters $\alpha,\beta,\sigma$ are determined by the central charge and the parameters of the averaging function. 
Owing to the Heaviside function, the distribution is supported in $[-\sigma,\infty)$
where $-\sigma$ is known on general grounds~\cite{FewFordRom} to equal the optimal quantum energy inequality (QEI) bound~\cite{FewHol1} that constrains the expectation value of $T(f)$ in (precisely delineated)  general states, 
\begin{equation}\label{eq:QEI}
        \langle T(f)\rangle_\psi 
                            \geq -\sigma=
                                    -\frac{c}{12\pi} \int \bigg{(} \frac{d}{du} \sqrt{f(u)} \bigg{)}^2 du .
\end{equation}  
The other main features of the distribution function~\eqref{eq:shiftedGamma} are the exponential tail and the power-law behaviour as $\omega\to -\sigma^+$, which is an integrable singularity if $\alpha\in (0,1)$ and regular for $\alpha\ge 1$.

We proceed as follows: in Section~\ref{probvac} we briefly describe the method of~\cite{FewFordRom}, then turning to the families~\eqref{eq:mGaussian} and~\eqref{eq:g} in Sections~\ref{sec:mGaussian} and~\ref{sec:mLorentz} respectively. An indirect argument is given in Section~\ref{sec:half} to read off the distributions for half-line variants. 

Given the probability distribution for a single chiral component of the stress-energy tensor, the distribution for timelike and spacetime averages of the full stress-energy tensor may be computed easily. We refer to~\cite{FewFordRom,FewFord:2019} for discussion. 

\section{Moment generating functions} \label{probvac}

We begin by briefly summarising the method of \cite{FewFordRom}. With $T(f)$ as above, the $n$'th moment of $T(f)$ in the vacuum state is
\begin{equation}
    \mathcal{G}_n[f] :=  \braket{\Omega|T(f)^n \Omega},
\end{equation}
and the moment generating function is defined by 
\begin{equation} \label{moment}
    M[\mu f] = \sum_{n=0}^{\infty} \frac{\mu^n \mathcal{G}_n [f]}{n!},
\end{equation} 
and the connected moment generating function is 
\begin{equation}
W[\mu f]= \log M[\mu f].
\end{equation}

In any CFT one has $\mathcal{G}_0[f]=1$ and $\mathcal{G}_1[f]=0$. Values of $\mathcal{G}_n[f]$ for $n\ge 2$ may be determined using the Ward identities in CFT. In particular, one has
\begin{equation}
\begin{split}
    \mathcal{G}_2 [f] &=
    \frac{c}{8\pi^2}\int_{\mathbb{R}^2}\frac{f(u_1)f(u_2)}{(u_2-u_1-i0)^4}du_1\,du_2 \\
    &=
    \frac{c}{48\pi^2}\int_0^\infty \omega^3 |\hat{f} (\omega)|^2 d\omega,
    \end{split}
\end{equation} 
where $\hat{f}$ is the Fourier transform
\begin{equation}
    \hat{f}(\omega) = \int_{-\infty}^\infty f(t) e^{-i\omega t}\,dt .
\end{equation} 
In~\cite{FewFordRom} (see also~\cite{Haba}) it was shown that the Ward identities
 allow one to write the connected moment generating function as
\begin{equation}
    W[\mu f]=  \int_0^\mu (\mu-\lambda) \mathcal{G}_2[f_\lambda] d\lambda,
\end{equation}
where $f_\lambda$ is a $1$-parameter family of functions solving the \emph{flow equation} 
\begin{equation} \label{eq:flow}
    \frac{df_\lambda}{d\lambda} = f_\lambda \star f_\lambda, \quad f_0=f
\end{equation}
where 
\begin{equation} \label{star}
        (f\star f)(u) = \int \frac{f(w) f'(u) - f'(w)f(u)}{2 \pi (w - u)} dw.
\end{equation}
Eq.~\eqref{eq:flow} is a nonlinear integro-differential equation. The only known closed form solutions require $f_0$ to be a Gaussian~\cite{FewFordRom} or a member of one of two infinite families~\cite{FewHol2}. This paper will provide infinitely many further exact solutions.

Once the moment generating function is found, the probability density function is obtained as an inverse Laplace transform, because
\begin{equation}
    M[\mu f] = \int_{-\infty}^\infty P(\omega) e^{\mu \omega} d\omega.
\end{equation}
This may be performed by inspection in certain cases. In particular, the shifted Gamma distribution \eqref{eq:shiftedGamma} has moment generating function
\begin{equation}\label{eq:MGFshiftedGamma}
   M(\mu) = \left( 1-\frac{\mu}{\beta}\right)^{-\alpha} e^{-\mu\sigma},
\end{equation}
and connected moment generating function
\begin{equation}\label{eq:cMGFshiftedGamma}
 W(\mu) = \alpha\log\left( \frac{\beta}{\beta-\mu}\right) -\mu\sigma .
\end{equation}
The moment generating function is of this form if and only if
\begin{equation}\label{eq:G2shiftedGamma}
    \mathcal{G}_2[f_\lambda] = \frac{\alpha}{(\beta-\lambda)^2}
\end{equation}
and $\sigma=\alpha/\beta$. 

It is important to note that the moments of a shifted Gamma distribution grow sufficiently slowly that the moment generating function uniquely determines the probability distribution, as a consequence of the Hamburger moment theorem. See~\cite{FewFordRom} for discussion and references.

\section{Modified Gaussian functions}\label{sec:mGaussian}

We now apply the method described  Sec.~\ref{probvac} to the modified Gaussian functions~\eqref{eq:mGaussian}, fixing the normalisation parameter $\gamma$ so that
$f_a$ has unit integral,  
\begin{equation}\label{eq:gamma}
    \gamma = \frac{b^{a+\frac{1}{2}}}{\Gamma \big{(}a+\frac{1}{2} \big{)}}.
\end{equation}
For the most part of this section we fix $a\in\NN_0$ and $b>0$ and drop the subscripts from $f_{a,b}$ to lighten the notation. 

Starting with the derivative  
\begin{equation}
        f'(x) 
                =  2\gamma x^{2a} e^{-bx^2} 
                    \bigg{(} \frac{a-bx^2}{x} \bigg{)}
                =  
                    2\bigg{(} \frac{a-bx^2}{x} \bigg{)} f(x),
\end{equation}
we calculate $(f\star f)(u)$, beginning with the integrand 
\begin{equation}
        \frac{f'(u)f(w)-f'(w)f(u)}{2\pi(w-u)} 
                                                = 
                                                    \frac{a+buw}{\pi uw} f(u) f(w).
\end{equation}
Noting that the term proportional to $a$ is odd in $w$, integration over $\mathbb{R}$ yields
\begin{equation} 
        (f\star f)(u) 
                        = 
                            \frac{u^{2a} e^{-bu^2} b \gamma^2}{\pi}
                            \int_{-\infty}^\infty 
                            w^{2a} e^{-bw^2} dw
                        =
                            \frac{b f(u)}{\pi},
\end{equation}
where we have used the definition of $f$ and the fact that it has unit integral. To solve the flow equation~\eqref{eq:flow} we now make the ansatz  
\begin{equation}
    f_\lambda (u) = A(\lambda) f(u),
\end{equation}
as used in~\cite{FewFordRom}, whereupon the flow equation reduces to the differential equation
\begin{equation}
    \frac{dA(\lambda)}{d \lambda} = \frac{A(\lambda)^2 b}{\pi}, \qquad A(0)=1.
\end{equation} 
Solving, we find
\begin{equation}
    A(\lambda) = \frac{\pi}{\pi-b \lambda}.
\end{equation}

Now that we have solved the flow equation, $W[\mu f]$ may be calculated easily because
$\mathcal{G}_2[f_\lambda]=A(\lambda)^2 \mathcal{G}_2[f]$ 
has the general form~\eqref{eq:G2shiftedGamma}. Restoring the subscripts $a,b$, the moment generating function is a shifted Gamma distribution with parameters 
\begin{equation}\label{eq:modGpars}
\alpha_{a,b} = \frac{\pi^2\mathcal{G}_2[f_{a,b}]}{b^2}, \qquad \beta_{a,b}=\frac{\pi}{b},\qquad  \sigma_{a,b}=\frac{\alpha}{\beta}.
\end{equation}

It remains to compute $\mathcal{G}_2 [f_{a,b}]$, for which we first compute the Fourier transform  
\begin{equation}
    \hat{f}_{a,b} (\omega) 
                        =
                            (-1)^a\frac{e^{-\omega^2/(4b)} \sqrt{\pi} }{2^{2a} \Gamma \big{(}a+\frac{1}{2} \big{)}} H_{2a}\left(\frac{\omega}{2\sqrt{b}}\right),
\end{equation}
using \cite[\S 18.10.10]{NIST:DLMF} and the fact that the Hermite polynomial $H_{2a}$ is even.
With this we then obtain
\begin{equation}
        \mathcal{G}_2 [f_{a,b}] 
                         = 
                            \frac{c\int_0^\infty \omega^3 e^{-\frac{\omega^2}{2b}}  H_{2a}\left(\frac{\omega}{2\sqrt{b}}\right)^2 d\omega}{3\cdot 2^{4(a+1)}\pi \Gamma \big{(}a+\frac{1}{2} \big{)}^2} .
\end{equation}
To evaluate this expression, we use a generating function argument, using the formula 
\begin{equation}\label{eq:Hermite}
e^{-2z^2}\sum_{n=0}^\infty \frac{(x/2)^n}{n!} H_n(z)^2 = \frac{e^{-2z^2/(1+x)}}{\sqrt{1-x^2}},
\end{equation}
which follows from \cite[\S 10.13(22)]{Bateman:vol2}, and implies
\begin{equation}
\begin{split}
 \lefteqn{\sum_{n=0}^\infty\frac{(x/2)^n}{n!}   \intno \omega^3 e^{-\frac{\omega^2}{2b}}  H_{n}\left(\frac{\omega}{2\sqrt{b}}\right)^2 d\omega} \\ &= \frac{2b^2(1+x)^2}{\sqrt{1-x^2}} \\
&= 2b^2 \left( 1+2x + x^2 \right) \suminf \frac{(2k-1)!!}{2^k k!} x^{2k}.
\end{split}
\end{equation}
Comparing coefficients of $x^{2a}$, we eventually reach the final expression
\begin{equation}
    \mathcal{G}_2 [f_{a,b}] = \frac{c b^2 (4a-1)}{24\pi^2 (2a-1)}.
\end{equation}
 
Using~\eqref{eq:modGpars}, we conclude that the probability distribution of measurements of $T(f_{a,b})$ in the vacuum state has a shifted Gamma probability density function~\eqref{eq:shiftedGamma} with parameters 
\begin{equation}
        \alpha_{a,b} = \frac{c(4a-1)}{24 (2a-1)}, 
       \quad
        \beta_{a,b} = \frac{\pi}{b  },
        \quad
        \sigma_{a,b} = \frac{c b (4a-1)}{24\pi (2a-1) }.
\end{equation}
As a consistency check, the value of $\sigma_{a,b}$, the lower bound of the distribution, can then be compared against the optimal QEI bound~\eqref{eq:QEI}.  
Noting that 
\begin{equation}
    \frac{d}{dv} \sqrt{f_{a,b}(v)} 
                                = 
                                    \sqrt{\gamma} v^{a-1} e^{-\frac{b}{2}v^2} (a-bv^2),
\end{equation}
consistency requires that
\begin{equation}
    \sigma_{a,b} = \frac{c \gamma }{6\pi} \int_{0}^{\infty} v^{2a-2} e^{-bv^2} (a-bv^2)^2 dv,
\end{equation}
and a routine calculation with Gamma functions confirms that this is indeed the case, as it should be on general grounds~\cite{FewFordRom}. A further consistency check is that the $a=0$ special case reproduces the example given in~\cite{FewFordRom}.

\section{Modified Lorentzian}\label{sec:mLorentz}

Our second family of test functions is the modified Lorentzian functions given in~\eqref{eq:g}. Proceeding as before, we suppress the subscripts $a,b$ and $n$ for the most part. 
We also make an ansatz similar to that used in~\cite{FewHol2}, defining $g_\lambda$ by replacing $b$ by $b_\lambda$ in \eqref{eq:g}, with $b_0=b$, and leaving $C$ fixed. 
The derivative of this test function is 
\begin{equation}
    g_\lambda'(u) = \frac{2Cu^{2a-1}}{(b_\lambda^2+u^2)^{n+1}} \Big{(} a b_\lambda^2+u^2(a-n)\Big{)}
\end{equation}
and a calculation gives
\begin{equation}
\begin{split}
    \lefteqn{\frac{g_\lambda'(u)g_\lambda(w)-g_\lambda'(w)g_\lambda(u)}{w-u}}\\ 
    &= \frac{2n(Cb_\lambda)^2(uw)^{2a}}{(b_\lambda^2+u^2)^{n+1}(b_\lambda^2+w^2)^{n+1}}  +\text{terms odd in $w$.}
\end{split}
\end{equation}
Integrating and dividing by $(2\pi)$,
\begin{equation} 
         (g_\lambda \star g_\lambda)(u)  = \frac{\tilde{C} n b_\lambda^{2(a-n)+1}g_\lambda(u)}{b_\lambda^2+u^2} ,
\end{equation}
where
\begin{equation}
    \tilde{C}=  \frac{C}{\pi} \int_{-\infty}^\infty \frac{x^{2a}}{(1+x^2)^{n+1}}dx =  \frac{C}{\pi}B\left(a+\tfrac{1}{2},n-a-\tfrac{1}{2}\right)
\end{equation} 
on using identity \cite[3.251]{GradshteynRyzhik:5thed} for the beta function. The same identity may be used to determine the value of $C$ so that $g_0$ has unit integral, namely 
\begin{equation}\label{eq:C}
   C = \frac{b_0^{2n-2a-1}}{B (a+\frac{1}{2}, n-a-\frac{1}{2})},
\end{equation}
in which case
\begin{equation}
    \tilde{C} = \frac{b_0^{2(n-a)-1}(2(n-a)-1)}{2n\pi}
\end{equation}
on simplifying the ratio of Beta functions. 

On the other hand, the derivative with respect to $\lambda$ is
\begin{equation} 
         \frac{dg_\lambda (u)}{d \lambda}  
                                        =
                                            -\frac{2n b_\lambda g_\lambda(u)}{b_\lambda^2+u^2}
                                            \frac{db_\lambda}{d\lambda},
\end{equation}  
so the flow equation reduces to  
\begin{equation} 
        b_\lambda^{2(n-a)}\frac{db_\lambda}{d\lambda}
       =-\frac{\tilde{C}}{2}=-\frac{b_0^{2(n-a)-1}(2(n-a)-1)}{4n\pi},
\end{equation}
which has unique solution  
\begin{equation}
     b_\lambda
                    =
                        b_0 \bigg{(} 1 - \frac{4(n-a)^2-1}{4n\pi b_0^{2}}\lambda \bigg{)}^{\frac{1}{2n-2a+1}},
\end{equation}
completing the solution of the flow equation.

The next step is to compute $\mathcal{G}_2[g_\lambda]$, facilitated by the observation that
\begin{equation}
    g_\lambda(u) = (b_\lambda/b_0)^{2(a-n)} g_0(u b_0/b_\lambda),
\end{equation}
giving 
\begin{equation}
\begin{split}
    \mathcal{G}_2[g_\lambda] &= \mathcal{G}_2[g_0](b_\lambda/b_0)^{4(a-n)-2} \\
    &= 
    \mathcal{G}_2[g_0]\left(\frac{4n\pi b_0^2}{4n\pi b_0^2- (4(n-a)^2-1)\lambda}\right)^2,
\end{split}
\end{equation}
which is of the form~\eqref{eq:G2shiftedGamma}.

Consequently, it is clear that the probability distribution will again be a shifted Gamma distribution. One of the three parameters is already known, namely (on reinstating the $n,a,b$ subscripts) 
\begin{equation}\label{eq:betana}
    \beta_{n,a,b} = \frac{4n\pi b^{2}}{4(n-a)^2-1}
\end{equation}
while $\alpha_{n,a,b} = \mathcal{G}_2[g_{n,a,b}] \beta^2$ and $\sigma_{n,a,b} = \alpha_{n,a,b}/\beta_{n,a,b}$.
Rather than computing $\mathcal{G}_2[g_{n,a,b}]$ directly, we obtain the parameter $\sigma_{n,a,b}$ from the optimal QEI bound. This given by  
\begin{equation}\label{eq:sigmana}
    \begin{split}
        \sigma_{n,a,b}  &=
                                    \frac{cC}{12\pi} \int_{-\infty}^{\infty} \bigg{(} \frac{d}{dv}\sqrt{g_{n,a}(v)}\bigg{)}^2 dv, \\
                            &=
                                    \frac{cC}{12\pi} \int_{0}^{\infty} v^{a-\frac{3}{2}}(av-nv+a)^2(v+1)^{-n-2} dv,\\ 
                            &=
                                    \frac{c(1-2n+2a)(4a^2-4an-4a+n)}{48 \pi (2a-1)(n+1)b^2}.
    \end{split}
\end{equation}
Thus
\begin{equation}\label{eq:alphana}
    \alpha_{n,a,b} = \frac{cn(4a^2-4an-4a+n)}{12(2a-2n-1)(n+1)(2a-1)}.
\end{equation}
Equations~\eqref{eq:betana},~\eqref{eq:sigmana} and~\eqref{eq:alphana} give the three parameters of the shifted Gamma distribution for this family of test functions. Again, the special case $a=0$ reduces to a known example~\cite{FewHol2}. 

In passing we note that consistency between the two formulae for $\sigma_{n,a,b}$ (and then setting $b=1$) indirectly proves the identity, 
\begin{equation}\label{eq:identity}
\begin{split}
    \lefteqn{\int_0^\infty k^3 \left(\mathcal{F}[u^{2a}/(1+u^2)^n](k)\right)^2 \,dk} \\&= -\frac{(4(n-a)^2-1)^2(4a^2-4an-4a+n)}{4(2(n-a)+1)n(n+1)(2a-1)}\\
&\qquad \times    B\left(a+\tfrac{1}{2},n-a-\tfrac{1}{2}\right)^2,
\end{split}    
\end{equation}
where $\mathcal{F}$ denotes Fourier transformation
and in this case is (cf.~\cite[7.12(27)]{Bateman:vol2} for the $a=0$ case)
\begin{equation}
\begin{split}
    \lefteqn{\mathcal{F}[u^{2a}/(1+u^2)^n](k)} \\&= (-1)^a\frac{2\sqrt{\pi}}{(n-1)!}\frac{d^{2a}}{dk^{2a}} (k/2)^{n-1/2} K_{n-1/2}(k),
\end{split}
\end{equation}
where $K_\nu$ denotes a modified Bessel function. We have no independent proof for the identity~\eqref{eq:identity}, but have tested it in a number of cases (all $0\le a<n$, $1\le n\le 10$). The case $a=0$ was implicitly established in~\cite{FewHol2} for arbitrary $n\ge 1$.

\section{Half-sided variants}\label{sec:half}

When the parameter $a$ is at least $1$, both $f_{a,b}$ and $g_{n,a,b}$ vanish to at least quadratic order at the origin. As we will show, this is sufficient for the two halves of the real line to decouple. Consequently we find a simple derivation for the probability distribution of half-line versions of these averaging functions.

Our argument is based on the Virasoro relations, which imply that
\begin{equation}\label{eq:Virasoro}
\begin{split}
    -i[T(g),T(h)] &= T(gh'-g'h) \\ &\quad + \frac{c}{24\pi}\int_{-\infty}^\infty g'''(u)h(u)\,du.
\end{split}
\end{equation}
Let $f$ be one of the test functions studied above, with $a\ge 1$, and define
\begin{equation}
    g(u) = \Theta(u) f(u),\qquad h(u) = \Theta(-u) f(u).
\end{equation}
Then one may check that both $gh'-g'h$ and $g'''h$ vanish identically, so that $T(f)$ may be expressed as the sum of two commuting operators (see the end of this section for a remark on the validity of this step).
Therefore the probability distribution associated with $T(f)$ should be the convolution of the distributions associated with $T(g)$ and $T(h)$. But as $h$ is simply the reflection of $g$ in the origin, these distributions are the same. We conclude that
the probability density function for $f$ is the convolution square of the probability density function for $g$, or conversely, that the probability density function for $g$ is a convolution square root of that for $f$. As the shifted Gamma distribution is (infinitely) divisible, this implies that the distribution for $g$ is again a shifted Gamma distribution, with parameters $\alpha$ and $\sigma$ divided by two and $\beta$ unchanged.

Summarising, we have argued that energy density average against the family 
\begin{equation}\label{eq:halfmGaussian}
\tilde{f}_{a,b}(u) =  \gamma \Theta(u) u^{2a} e^{-bu^2},\qquad 
    a\in \mathbb{N}_{>0}, \quad b>0,
\end{equation}
where $\gamma$ is the same normalisation factor as in Eq.~\eqref{eq:gamma} (so $\tilde{f}_{a,b}$ has total integral $\tfrac{1}{2}$) correspond to shifted Gamma distributions with parameters
\begin{equation}
        \tilde{\alpha}_{a,b} = \frac{c(4a-1)}{48 (2a-1)}, 
       \quad
        \tilde{\beta}_{a,b} = \frac{\pi}{b  },
        \quad
        \tilde{\sigma}_{a,b} = \frac{c b (4a-1)}{48\pi (2a-1) }.
\end{equation}
Similarly, averages against the family  
\begin{equation}\label{eq:halfg}
    g_{n,a,b}(u) = \frac{C\Theta(u) u^{2a}}{(b^2+u^2)^n}, \quad a,n\in\mathbb{N},~1\le a< n,~b>0,
\end{equation}
with $C$ as in Eq.~\eqref{eq:C}, correspond to shifted Gamma distributions with parameters 
\begin{equation}
\begin{split}
    \tilde{\alpha}_{n,a,b} &= \frac{cn(4a^2-4an-4a+n)}{24(2a-2n-1)(n+1)(2a-1)} \\
    \tilde{\beta}_{n,a,b} &= \frac{4n\pi b^{2}}{4(n-a)^2-1} \\
    \tilde{\sigma}_{n,a,b}&= \frac{c(1-2n+2a)(4a^2-4an-4a+n)}{96 \pi (2a-1)(n+1)b^2}.
\end{split}
\end{equation}
It would be interesting to confirm these results by direct solutions of the flow equation, but we do not pursue this here. 

To close we note that our argument depended on an extension of the Virasoro relations~\eqref{eq:Virasoro} to non-smooth test functions, so some care is needed. The argument is rigorously valid for $a\ge 3$ at least, because then $g,h$ are at least four times continuously differentiable (see~\cite{CarpiWeiner:2005} for rigorous proofs; more recent work in a similar direction is presented in~\cite{DelVecchio2019,CarpiDelVecIovTan:2018}). It is a reasonable conjecture that the overall result holds for $a=2$ (and perhaps even $a=1$ as well) and this is supported by numerical evidence to be reported elsewhere.

\section{Concluding remarks}

We have described four new infinite families of averaging functions for which the probability distribution of the vacuum CFT energy density
may be computed in closed form. In all these examples -- like those previously obtained in~\cite{FewFordRom,FewHol2} by the moment generating function method -- the result is a shifted Gamma distribution. While it is known that this cannot be the distribution in all cases, these results suggest that there is a large distinguished class of averaging functions for which this is the case. Understanding this may help to cast light on the problem of finding general solutions to the flow equation, and related issues, such as the relationship between the moment generating function method and the conformal welding method, which is currently unclear. 
 
\begin{acknowledgments}
We thank Atsushi Higuchi for directing us towards the generating function for Hermite polynomials used in~\eqref{eq:Hermite}. M.C.A. thanks the Department of Mathematics for financial support during the final phases of this work, and C.J.F. thanks Sebastiano Carpi for a useful conversation relating to Refs.~\cite{CarpiWeiner:2005,DelVecchio2019,CarpiDelVecIovTan:2018} and Bradley Lang for conducting the numerical calculations mentioned in Sec.~\ref{sec:half}.
\end{acknowledgments}


\begin{thebibliography}{16}
	\expandafter\ifx\csname natexlab\endcsname\relax\def\natexlab#1{#1}\fi
	\expandafter\ifx\csname bibnamefont\endcsname\relax
	\def\bibnamefont#1{#1}\fi
	\expandafter\ifx\csname bibfnamefont\endcsname\relax
	\def\bibfnamefont#1{#1}\fi
	\expandafter\ifx\csname citenamefont\endcsname\relax
	\def\citenamefont#1{#1}\fi
	\expandafter\ifx\csname url\endcsname\relax
	\def\url#1{\texttt{#1}}\fi
	\expandafter\ifx\csname urlprefix\endcsname\relax\def\urlprefix{URL }\fi
	\providecommand{\bibinfo}[2]{#2}
	\providecommand{\eprint}[2][]{\url{#2}}
	
	\bibitem[{\citenamefont{{Huang} and {Ford}}(2017)}]{HuangFord:2017}
	\bibinfo{author}{\bibfnamefont{H.}~\bibnamefont{{Huang}}} \bibnamefont{and}
	\bibinfo{author}{\bibfnamefont{L.~H.} \bibnamefont{{Ford}}},
	\bibinfo{journal}{\prd} \textbf{\bibinfo{volume}{96}}, \bibinfo{eid}{016003}
	(\bibinfo{year}{2017}), \eprint{1610.01252}.
	
	\bibitem[{\citenamefont{Fewster et~al.}(2010)\citenamefont{Fewster, Ford, and
			Roman}}]{FewFordRom}
	\bibinfo{author}{\bibfnamefont{C.~J.} \bibnamefont{Fewster}},
	\bibinfo{author}{\bibfnamefont{L.~H.} \bibnamefont{Ford}}, \bibnamefont{and}
	\bibinfo{author}{\bibfnamefont{T.~A.} \bibnamefont{Roman}},
	\bibinfo{journal}{Phys. Rev. D} \textbf{\bibinfo{volume}{81}},
	\bibinfo{pages}{{121}{901}} (\bibinfo{year}{2010}), \eprint{arXiv:1004.0179}.
	
	\bibitem[{\citenamefont{Fewster and Ford}(2015)}]{FewsFord:2015}
	\bibinfo{author}{\bibfnamefont{C.~J.} \bibnamefont{Fewster}} \bibnamefont{and}
	\bibinfo{author}{\bibfnamefont{L.~H.} \bibnamefont{Ford}},
	\bibinfo{journal}{Phys. Rev. D} \textbf{\bibinfo{volume}{92}},
	\bibinfo{pages}{{105}{008}} (\bibinfo{year}{2015}),
	\urlprefix\url{http://link.aps.org/doi/10.1103/PhysRevD.92.105008}.
	
	\bibitem[{\citenamefont{Fewster et~al.}(2012)\citenamefont{Fewster, Ford, and
			Roman}}]{FewForRom:2012}
	\bibinfo{author}{\bibfnamefont{C.~J.} \bibnamefont{Fewster}},
	\bibinfo{author}{\bibfnamefont{L.~H.} \bibnamefont{Ford}}, \bibnamefont{and}
	\bibinfo{author}{\bibfnamefont{T.~A.} \bibnamefont{Roman}},
	\bibinfo{journal}{Phys. Rev. D} \textbf{\bibinfo{volume}{85}},
	\bibinfo{pages}{{125}{038}} (\bibinfo{year}{2012}), \eprint{1204.3570}.
	
	\bibitem[{\citenamefont{Schiappacasse et~al.}(2018)\citenamefont{Schiappacasse,
			Fewster, and Ford}}]{SchiaFewsFord:2018}
	\bibinfo{author}{\bibfnamefont{E.~D.} \bibnamefont{Schiappacasse}},
	\bibinfo{author}{\bibfnamefont{C.~J.} \bibnamefont{Fewster}},
	\bibnamefont{and} \bibinfo{author}{\bibfnamefont{L.~H.} \bibnamefont{Ford}},
	\bibinfo{journal}{Phys. Rev. D} \textbf{\bibinfo{volume}{97}},
	\bibinfo{pages}{025013} (\bibinfo{year}{2018}),
	\urlprefix\url{https://link.aps.org/doi/10.1103/PhysRevD.97.025013}.
	
	\bibitem[{\citenamefont{Fewster and Hollands}(2019)}]{FewHol2}
	\bibinfo{author}{\bibfnamefont{C.~J.} \bibnamefont{Fewster}} \bibnamefont{and}
	\bibinfo{author}{\bibfnamefont{S.}~\bibnamefont{Hollands}},
	\bibinfo{journal}{Lett. Math. Phys.} \textbf{\bibinfo{volume}{109}},
	\bibinfo{pages}{747} (\bibinfo{year}{2019}),
	\urlprefix\url{https://doi.org/10.1007/s11005-018-1124-6}.
	
	\bibitem[{\citenamefont{{Gawedzki} and {Kozlowski}}(2019)}]{GawedKozl:2019}
	\bibinfo{author}{\bibfnamefont{K.}~\bibnamefont{{Gawedzki}}} \bibnamefont{and}
	\bibinfo{author}{\bibfnamefont{K.~K.} \bibnamefont{{Kozlowski}}},
	\bibinfo{journal}{arXiv e-prints} \bibinfo{eid}{arXiv:1906.04276}
	(\bibinfo{year}{2019}), \eprint{1906.04276}.
	
	\bibitem[{\citenamefont{Haba}(1990)}]{Haba}
	\bibinfo{author}{\bibfnamefont{Z.}~\bibnamefont{Haba}}, \bibinfo{journal}{Phys.
		Rev. D} \textbf{\bibinfo{volume}{41}}, \bibinfo{pages}{724}
	(\bibinfo{year}{1990}),
	\urlprefix\url{http://dx.doi.org/10.1103/PhysRevD.41.724}.
	
	\bibitem[{\citenamefont{Fewster and Hollands}(2005)}]{FewHol1}
	\bibinfo{author}{\bibfnamefont{C.~J.} \bibnamefont{Fewster}} \bibnamefont{and}
	\bibinfo{author}{\bibfnamefont{S.}~\bibnamefont{Hollands}},
	\bibinfo{journal}{Rev. Math. Phys.} \textbf{\bibinfo{volume}{17}},
	\bibinfo{pages}{577} (\bibinfo{year}{2005}),
	\urlprefix\url{https://doi.org/10.1142/S0129055X05002406}.
	
	\bibitem[{\citenamefont{Fewster and Ford}(2019)}]{FewFord:2019}
	\bibinfo{author}{\bibfnamefont{C.~J.} \bibnamefont{Fewster}} \bibnamefont{and}
	\bibinfo{author}{\bibfnamefont{L.~H.} \bibnamefont{Ford}}
	(\bibinfo{year}{2019}), \bibinfo{note}{in preparation}.
	
	\bibitem[{{\relax DLMF}()}]{NIST:DLMF}
	{\relax DLMF}, \emph{\bibinfo{title}{{\it NIST Digital Library of Mathematical
				Functions}}}, \bibinfo{howpublished}{http://dlmf.nist.gov/, Release 1.0.23 of
		2019-06-15}, \bibinfo{note}{{F}.~W.~J. Olver, A.~B. {Olde Daalhuis}, D.~W.
		Lozier, B.~I. Schneider, R.~F. Boisvert, C.~W. Clark, B.~R. Miller and B.~V.
		Saunders, eds.}, \urlprefix\url{http://dlmf.nist.gov/}.
	
	\bibitem[{\citenamefont{Erd\'{e}lyi et~al.}(1953)\citenamefont{Erd\'{e}lyi,
			Magnus, Oberhettinger, and Tricomi}}]{Bateman:vol2}
	\bibinfo{author}{\bibfnamefont{A.}~\bibnamefont{Erd\'{e}lyi}},
	\bibinfo{author}{\bibfnamefont{W.}~\bibnamefont{Magnus}},
	\bibinfo{author}{\bibfnamefont{F.}~\bibnamefont{Oberhettinger}},
	\bibnamefont{and} \bibinfo{author}{\bibfnamefont{F.~G.}
		\bibnamefont{Tricomi}}, \emph{\bibinfo{title}{Higher transcendental
			functions. {V}ol. {II}}} (\bibinfo{publisher}{McGraw-Hill Book Company, Inc.,
		New York-Toronto-London}, \bibinfo{year}{1953}), \bibinfo{note}{based, in
		part, on notes left by Harry Bateman}.
	
	\bibitem[{\citenamefont{Gradshteyn and Ryzhik}(1994)}]{GradshteynRyzhik:5thed}
	\bibinfo{author}{\bibfnamefont{I.~S.} \bibnamefont{Gradshteyn}}
	\bibnamefont{and} \bibinfo{author}{\bibfnamefont{I.~M.}
		\bibnamefont{Ryzhik}}, \emph{\bibinfo{title}{Table of integrals, series, and
			products}} (\bibinfo{publisher}{Academic Press, Inc., Boston, MA},
	\bibinfo{year}{1994}), \bibinfo{edition}{5th} ed., ISBN
	\bibinfo{isbn}{0-12-294755-X}, \bibinfo{note}{translation edited and with a
		preface by Alan Jeffrey}.
	
	\bibitem[{\citenamefont{Carpi and Weiner}(2005)}]{CarpiWeiner:2005}
	\bibinfo{author}{\bibfnamefont{S.}~\bibnamefont{Carpi}} \bibnamefont{and}
	\bibinfo{author}{\bibfnamefont{M.}~\bibnamefont{Weiner}},
	\bibinfo{journal}{Comm. Math. Phys.} \textbf{\bibinfo{volume}{258}},
	\bibinfo{pages}{203} (\bibinfo{year}{2005}),
	\urlprefix\url{https://doi.org/10.1007/s00220-005-1335-4}.
	
	\bibitem[{\citenamefont{Del~Vecchio et~al.}(2019)\citenamefont{Del~Vecchio,
			Iovieno, and Tanimoto}}]{DelVecchio2019}
	\bibinfo{author}{\bibfnamefont{S.}~\bibnamefont{Del~Vecchio}},
	\bibinfo{author}{\bibfnamefont{S.}~\bibnamefont{Iovieno}}, \bibnamefont{and}
	\bibinfo{author}{\bibfnamefont{Y.}~\bibnamefont{Tanimoto}},
	\bibinfo{journal}{Communications in Mathematical Physics}
	(\bibinfo{year}{2019}),
	\urlprefix\url{https://doi.org/10.1007/s00220-019-03419-2}.
	
	\bibitem[{\citenamefont{{Carpi} et~al.}(2018)\citenamefont{{Carpi}, {Del
				Vecchio}, {Iovieno}, and {Tanimoto}}}]{CarpiDelVecIovTan:2018}
	\bibinfo{author}{\bibfnamefont{S.}~\bibnamefont{{Carpi}}},
	\bibinfo{author}{\bibfnamefont{S.}~\bibnamefont{{Del Vecchio}}},
	\bibinfo{author}{\bibfnamefont{S.}~\bibnamefont{{Iovieno}}},
	\bibnamefont{and}
	\bibinfo{author}{\bibfnamefont{Y.}~\bibnamefont{{Tanimoto}}},
	\bibinfo{journal}{arXiv e-prints} \bibinfo{eid}{arXiv:1808.02384}
	(\bibinfo{year}{2018}), \eprint{1808.02384}.
	
\end{thebibliography}
\end{document}